# Room Temperature Optically and Magnetically Active Edges in Phosphorene Nanoribbons

[1]Arjun Ashoka, [2,3]Adam J. Clancy[#], [4]Naitik A. Panjwani[#], [5]Adam Cronin, [6]Loren Picco, [3]Eva S. Y. Aw, [1]Nicholas J. M. Popiel, [1]Alex Eaton, [7]Thomas G. Parton, [3]Rebecca R. C. Shutt, [1]Sascha Feldmann, [1]Remington Carey, [2,8,9]Thomas J. Macdonald, [10]Marion E. Severijnen, [10]Sandra Kleuskens, [11]Loreta A. Muscarella, [5,12,13]Felix R. Fischer, [14]Hilton Barbosa de Aguiar, [1]Richard H. Friend, [4]Jan Behrends, [10]Peter C.M. Christianen, [3]Christopher A. Howard, and [1,14]Raj Pandya

[1]Cavendish Laboratory, University of Cambridge, J.J. Thomson Avenue, CB3 0HE, Cambridge, United Kingdom
[2]Department of Chemistry, UCL, Christopher Ingold Building, Gordon St, London WC1H 0AJ, UK
[3]Department of Physics & Astronomy, University College London, London, WC1E 6BT, United Kingdom
[4]Berlin Joint EPR Lab, Fachbereich Physik, Freie Universität Berlin, D-14195 Berlin, Germany
[5]Department of Chemistry, University of California, Berkeley, Berkeley, California 94720, United States
[6]Department of Physics, Virginia Commonwealth University, Richmond, VA, USA
[7]Yusuf Hamied Department of Chemistry, University of Cambridge, Lensfield Road, Cambridge, CB2 1EW, United Kingdom
[8]Department of Chemistry and Centre for Processable Electronics, Imperial College London, London, W12 0BZ, United Kingdom
[9]Department of Electronic & Electrical Engineering, University College London, London WC1E 7JE, United Kingdom
[10]High Field Magnet Laboratory (HFML - EMFL), Radboud University, 6525 ED, Nijmegen, The Netherlands
[11]Center for Nanophotonics, AMOLF, Science Park 104, 1098 XG Amsterdam, The Netherlands
[12]Kavli Energy NanoSciences Institute at the University of California Berkeley and the Lawrence Berkeley National Laboratory, Berkeley, California 94720, United States
[13]Materials Sciences Division, Lawrence Berkeley National Laboratory, Berkeley, California 94720, United States
[14]Laboratoire Kastler Brossel, ENS-Université PSL, CNRS, Sorbonne Université, Collège de France, 24 rue Lhomond, 75005, Paris, France

#Denotes equal contribution
Correspondence: rp558@cam.ac.uk

**Nanoribbons – nanometer wide strips of a two-dimensional material – are a unique system in condensed matter. They combine the exotic electronic structures of low-dimensional materials with an enhanced number of exposed edges, where phenomena including ultralong spin coherence times[1–4], quantum confinement[5] and topologically protected states[6–8] can emerge. An exciting prospect for this material concept is the potential for both a tunable semiconducting electronic structure and magnetism along the nanoribbon edge. This combination of magnetism and semiconducting properties is the first step in unlocking spin-based electronics such as non-volatile transistors, a route to low-energy computing[9], and has thus far typically only been (simultaneously) observed in doped semiconductor systems and/or at low temperatures[10–15]. Here, we report the magnetic and semiconducting properties of phosphorene nanoribbons (PNRs). Static (SQUID) and dynamic (EPR) magnetization probes demonstrate that at room temperature, films of PNRs exhibit macroscopic magnetic properties, arising from their edge, with internal fields of ~240 to 850 mT. In solution, a giant magnetic anisotropy enables the alignment of PNRs at modest sub-1 T fields. By leveraging this alignment effect, we discover that upon photoexcitation, energy is rapidly funneled to a state that is**

**localized to the magnetic edge and coupled to a symmetry-forbidden edge phonon mode. Our results establish PNRs as a fascinating system for studying the interplay between magnetism and semiconducting ground states at room temperature and provide a stepping-stone towards using low-dimensional nanomaterials in quantum electronics.**

Over the last 40 years, strategies to couple different magnetic phenomena and the electronic ground states of semiconductors have mostly focussed on doping traditional semiconductors with certain transition metals to form so-called dilute magnetic semiconductors (DMSs)[16–20]. A more recent approach has been to exploit the unique spin configurations that can be realised in two-dimensional (2D) monolayer materials – some of which, such as CrSBr and $NiPS_3$, have been also shown to host (at low temperature) semiconducting magnetic phases[21–24]. Slicing two-dimensional monolayers into thin strips – nanoribbons – opens the possibility for an even wider range of magnetic spin-arrangements across and along the ribbon edges, whilst retaining the favourable electronic and mechanical properties of 2D systems. The most intensely studied nanoribbons thus far have been those based on the flat, honeycomb lattice of graphene[25,26] (GNRs), where at low temperatures, topologically-engineered bands[6,7,27] and a range of spin ordered electronic states have been observed[1,8]. However, the laborious chemical synthesis[28,29] combined with challenging edge-functionalisation[1,8] – akin to magnetic doping in DMSs – necessary to achieve the aforementioned phenomena, represents a significant challenge for both practical use and exploring fundamental nanoribbon properties in GNRs. Consequently, there is a need to explore nanoribbon systems which can be produced at scale and exhibit room-temperature intrinsic spintronic and electronic properties. Among the mooted systems, phosphorene nanoribbons (PNRs) – the black phosphorus analogue to GNRs – have uniquely been proposed.

From a synthesis standpoint, PNRs can be readily and scalably fabricated, including *via* top down approaches, that produce micron-length, high-aspect ratio ribbons with straight, uniform edges, and a long-axis aligned exclusively in the zigzag crystallographic direction of the bP parent lattice (**Figure 1a**)[30]. These are key characteristics for nanoribbon-based applications and are often challenging to simultaneously achieve in GNRs where the zigzag edge formation energy is large[31] and lengths are normally sub-100 nm. Although PNRs are expected to host many superlative characteristics for thermoelectric[31], transistor[32], and electrochemical[33], applications, the predicted presence of a visible band-gap[34] with high absorption (excitonic effects) and inherent edge-ferromagnetism, with a high Curie temperature, makes PNRs especially intriguiging[35–38]. Indeed, the highly anisotropic crystal and electronic structure of PNRs opens the possibility for a wide range of spin interactions both in- and out-of-the plane in single PNRs as well as PNR thin-films[39,40], and makes PNRs uniquely suited for studying and exploiting the intersection between functional electronic properties, external stimuli and magnetism in low-dimensions[9,41].

In this work we study PNRs with zigzag aligned edges[30] at the ensemble level in solution and deposited on substrates as well as individually on substrates (see Supplementary Note 1 and 2). PNRs are produced in batch solutions with average ribbon widths of ~15 nm, average lengths of ~700 nm, and predominantly as monolayers (>70%; see Methods for details on synthesis/characterisation and Supplementary Note 1 for batch-to-batch reproducibility)[38,3]. **Figure 1a** shows an Atomic Force Microscope (AFM) image of a typical individual monolayer ribbon (height 0.58 ± 0.14 nm and width 8.12 ± 1.08 nm) along with histograms of its width and height distribution measured along the ribbon using high-speed AFM (see Supplementary Note 3 for additional AFM details, discussion and images). Generally, along a given PNR the standard deviation in the width (width uniformity) is always AFM measurement-resolution limited *i.e.*, <1 nm.

Quantum Monte Carlo and dynamic mean field theory calculations have predicted that such PNRs as described above can host room temperature ferromagnetism on their edges[34,39,a]. More specifically, the PNR widths that we study are in the predicted crossover regime between decoherence-dominated classical magnetism and fluctuation-dominated quantum magnetism, in graphene nanoribbons[34,42,43]. PNRs in our ensemble samples are all found to be greater than the 3 nm width suggested to be required for stable edge magnetism[34].

To investigate the magnetic ground state of PNRs, it is important to first establish the degree of intrinsic magnetic anisotropy[44,45], due to their highly anisotropic shape and crystal structure. A non-invasive route to accessing the magnetic anisotropy of nano-objects dispersed in solution is to measure the magnetic field-induced alignment through the linear dichroism (LD) and linear birefringence (LB)[46]. The principle behind optically-detected magnetic alignment is that a magnetically-anisotropic nano-object orients in a magnetic field to minimise its magnetic orientational energy, which can eventually overcome thermal fluctuations. The optical anisotropy (LD or LB) of the underlying nano-object can then serve as a sensitive probe of the degree of alignment due to the external magnetic field. As shown in **Figure 1b**, we observe extremely unusual behaviour in the magnetic linear birefringence of PNRs, with the orientational anisotropy saturating at near 10 T at room temperature, much lower than that of other known nanomaterials[47–49]. On fitting the data to the orientational magnetic energy we retrieve the anisotropy in the volume magnetic susceptibility ($\Delta\chi$) multiplied by the volume of the material, $\Delta\chi V = 5 \times 10^{-27}\ \mathrm{m}^3$ (see Supplementary Note 6). We rule out tangling/bundling of PNRs in solution allowing us to use a rigid rod approximation when fitting the data in **Figure 1b** (see also Supplementary Note 7). We additionally observe a magnetic field induced LD signal at 543 nm, with more light absorbed by the PNR solution when the light is polarised along the magnetic field direction (**Figure 1b**, inset). PNRs are expected to have a higher absorption cross section with light polarised perpendicular to the zigzag direction, as has been reported experimentally for phosphorene[44] and obtained from GW-BSE calculation on PNRs[37,45]. We hence conclude that in a magnetic field, the ribbons orient with their short axis along the field direction, as an extremely strong out-of-plane polarized dipole is unlikely.

We then use reported[30] distributions of the length (940 values), width (940 values) and height (240 values) of PNRs, measured using TEM and AFM, to calculate an approximate distribution of all possible ribbon volumes in the solution[30] (see Supplementary Note 8 for a discussion of how the volume histogram is generated). We note as shown in the upper panel of **Figure 1c**, we find strong consistency in the length, width, and thickness distribution of PNRs between different batches used in this work, and also with dimension values measured in our original PNR synthesis work[30] (see also Supplementary Note 1). The above reproducibility in samples allows us to use and combine data from different PNR batches and link our ensemble data to PNR properties.

---

[a]In the theory literature that we reference in this work, the term 'zigzag PNR' refers to both the edge alignment and edge termination, unless otherwise stated. Experimentally, we can only conclusively state the PNRs we study are zigzag aligned. A discussion of the possible edge terminations/reconstructions present in our PNRs *via* polarised impulsive Raman spectroscopy, STM and theory calculations in provided in Supplementary Notes 2, 5, 6 and 23.

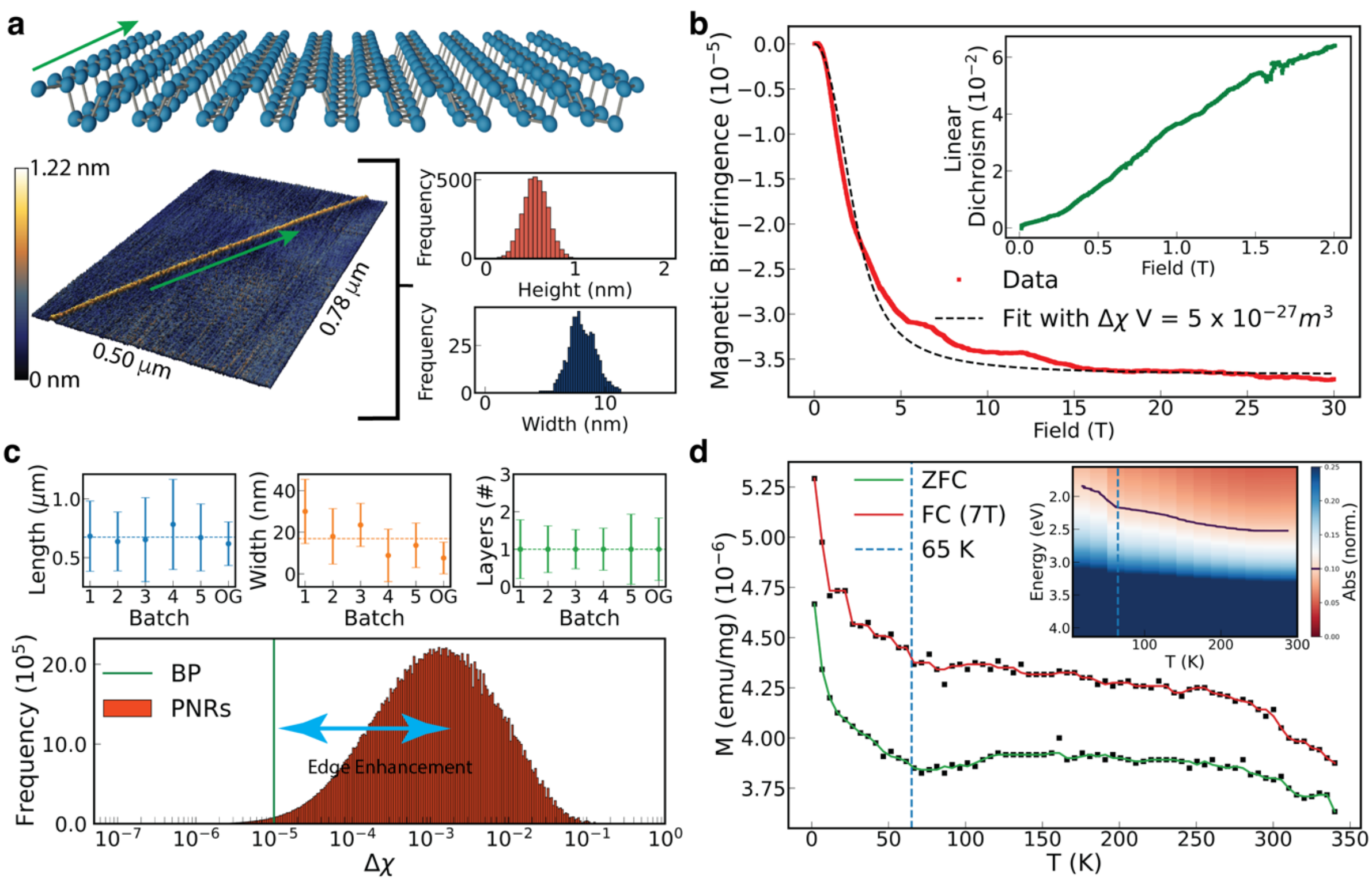

**Figure 1: Shape, magnetic and optical anisotropy of PNRs: a.** (Top) Crystal structure of the synthesised PNRs with the zigzag-aligned edges along their long axis (green arrow). (Bottom) Representative AFM image and histogram of height and width along a ribbon. An average width of 8 nm and average monolayer height of 0.6 nm is found along this ribbon. There are many more height pixels on top of the PNR than there are width slices resulting in a different sampling between the two quantities (see Methods and Supplementary Note 3 for extensive further discussion of the AFM measurement methods and abstraction). The height resolution for our AFM measurements is ± 15 pm, whereas X and Y lateral pixel sizes are 0.72 nm 1.24 nm, respectively. The standard deviation in width for all PNRs in this work is found to be <1 nm. **b.** Alignment of PNRs in solution measured using magnetic field induced linear birefringence up to 30 T shows a saturation at 10 T. Combined with the magnetic linear dichroism (inset), these results suggest that PNRs align with their short axis along the field direction. PNRs do not tangle/bundle in solution (see Supplementary Note 3 and 7) hence we fit the data using a rigid-rod approximation for the PNRs[50]. Data in red, fit in black **c.** (Bottom panel) Histogram of the intrinsic anisotropy calculated using the distribution of all possible PNR volumes in the solution based of TEM and AFM data (see Supplementary Note 1 and 8). The PNR magnetic susceptibility shows at least a 100-fold edge enhancement from the previously reported black phosphorous (BP) limit[44]. Top panel shows the mean (and standard deviation) in length, width and layer number across 5 PNR batches measured in this work, and that from the original synthesis reported by Watts *et al.*[30] ('OG'). There is an exceptional batch-to-batch reproducibility (see Supplementary Note 1). **d.** Field-cooled (FC) at 7 T and zero-field cooled (ZFC) magnetisation as a function of temperature for a batch PNR samples prepared in a plastic straw (which has a purely diamagnetic background (see Supplementary Note 10)) measured in a projecting field of 50 mT (see Figure S10.1 for projecting field dependence). The M *vs* T curve are quantitively identical for PNRs on a quartz rod (following raw SQUID background subtraction using the methods in ref[51]) and a similar response is observed for PNRs in frozen solution. The FC/ZFC curves are split at all temperatures, with a discontinuity around 65 K indicating a phase transition (see M *vs* 1/T plots in Figure S14.1). Inset shows PNR absorption spectrum between 300 and 10 K, here there is also a marked discontinuity near 65 K (the dark blue line indicated a contour at an absorption of 0.1 indicated in the colorbar).

From the distribution of PNR volumes in solution and the calculated $\Delta\chi V$, we extract an average anisotropy in the dimensionless magnetic susceptibility $\Delta\chi \approx 10^{-3}$ (**Figure 1c**, bottom). This is two orders of magnitude larger than the values from reports of $\Delta\chi$ for layered phosphorene, *i.e.*, black phosphorous[44] (**Figure 1c**, green line). The latter has its predominant magnetic anisotropy between the in- and out-of-plane directions, which are not expected to be the dominant anisotropy directions here[44]. We infer that this 100-fold enhancement in $\Delta\chi$ likely arises from the enhanced number of edges of the ribbons as this is the main intrinsic difference between the layered phosphorene and the PNRs. The origin of the large $\Delta\chi$ could be from diamagnetic, paramagnetic or ferromagnetic anisotropies, or a combination of these. It is therefore critical to establish the presence of unpaired spins in the PNR system, which would point to a para- or ferromagnetic origin for the large $\Delta\chi$, and the presence of the predicted edge magnetism.

In order to do this, we performed SQUID magnetometry on an ensemble of drop-cast PNRs in the solid state (see Supplementary Note 9 for a discussion on the morphology of all solid-state samples used in this work)[52]. We chose to drop cast the PNRs directly into a plastic straw to mitigate for any non-uniform background diamagnetic artefacts (see Figure S10.5 and S10.6 for SQUID measurements on empty plastic straws). The 7 T-field cooled (FC) and zero field cooled (ZFC) magnetisation measured in a 50 mT projecting field (**Figure 1d**) shows non-Curie behaviour as we approach temperatures as high as 350 K where the two curves approach each other but do not yet overlap. This is independently confirmed against background-subtracted, and subsequently fitted, raw SQUID traces measured on a quartz rod (see Supplementary Note 10 and 11 and methods in ref[51]). The response also does not change when measured in lower projecting fields of 20 mT (see Figure S10.1), indicating that the behaviour does not emerge due to an interaction with the measurement/projecting field of the SQUID, however higher projecting fields appear to suppress the magnetic phase likely due to the projecting field polarizing all the spins during the measurement. In the solution phase (NMP and DMF solvents), we also find that the FC and ZFC magnetisations are split up to the freezing point of the solvent (see Figure S10.2), above which the liquid dispersions are no longer a macroscopic magnetic system. The invariance of the PNR SQUID signals in the frozen solution phase and when deposited on both plastic and quartz substrates suggests that the magnetic phase we find is potentially robust against strain[37,39] in these environments (see Supplementary Note 10). We repeated the solution measurements on oxygenated PNRs by bubbling air through the solution in an attempt to chemically degrade the ribbon edge[30,53–55] and find a purely diamagnetic response[56,57] (see Figure S10.4 and Supplementary Note 12 for a discussion of how PNRs chemically degrade in air). Elemental analysis (inductively coupled plasma optical emission spectroscopy (ICP-OES)) of the same PNR samples used for SQUID determines less than 12 parts per billion of Fe, Co, Ni, Cr and Mn, ruling these out impurity metals as an origin of the signal (see Supplementary Note 13). Overall, this suggests that the magnetic response of PNRs arises, as theoretically predicted, from the ribbon edge with the persistence of the magnetic behaviour up to room temperature[34,39,40]. Isothermal magnetic field sweeps in the solution phase below the freezing temperature of the PNR solution (Figure S10.2) do indicate some level of hysteresis, however it remains unclear if standard hysteresis, as expected for bulk ferromagnets, is applicable in the case of edge magnetism (see additional Supplementary Note 10 and 11 for further discussion and data).

Between 55 to 70 K in the M(T) data, a discontinuity is observed in **Figure 1d** (seen more clearly when plotting M *vs* 1/T as shown in Figure S14.1) a signature of a magnetic phase transition. A similar discontinuity is also observed in the optical absorption at this temperature (**Figure 1d** inset) as well as in temperature dependent Raman and photoluminescence measurements (see Supplementary Note 14). Further, this signature of a phase transition is most clearly seen, at the same temperature, in the EPR data, discussed below (**Figure 2b**). These independent experimental observations confirm that the optically active species (*i.e.*, the PNRs) are the source of our magnetic signal, and offer a first hint that the electrons involved in the magnetic phase transition could also be involved in the optical excitations.

Next, we investigate the local spin correlations along the edge of PNRs by performing continuous wave electron paramagnetic resonance (cwEPR) measurements on a film formed on the inner wall of an EPR sample tube (see Supplementary Note 9 for a discussion on the morphology of the sample). Consistent with the hypothesis of unpaired electrons at the ribbon edge we observe a broad ≈12-19 mT wide $g$≈2 EPR signal at room temperature as shown in **Figure 2a, b**. To our surprise, a wider magnetic field scan (0-1400 mT) reveals much more intense and even broader ferromagnetic resonance (FMR) signals (**Figure 2a**).

The EPR signal also exhibits a strong temperature dependence, where the linewidth and $g$-factor values are non-monotonic, with a divergence of the linewidth and the disappearance of the signal on cooling around 50/60 K, which reappears at lower temperatures (**Figure 2c)**. As discussed above, this behaviour also appears in the SQUID magnetisation (**Figure 1d**), absorption (**Figure 1d inset)**, Raman, and photoluminescence spectra (at a similar temperature ±10 K), and is characteristic of a phase transition, however further work is required to describe its nature (see Supplementary Note 14). We suggest that the smooth temperature dependence of the $g$-factor from 300 to 65 K could signal a slow spin-canting in the out-of-plane direction due to antiferromagnetic (AF) correlations which lowers the $g$-factor from the fully out-of-plane polarised value of 2.14 towards the in-plane value of 2.00[58]. Finally, it is important to note that while the FMR and $g$≈2 EPR signals unequivocally arise from PNRs (distinct from backgrounds of the resonator or sample tubes), we find the visibility of the latter signal to be especially sensitive to the film preparation (see Supplementary Note 15).

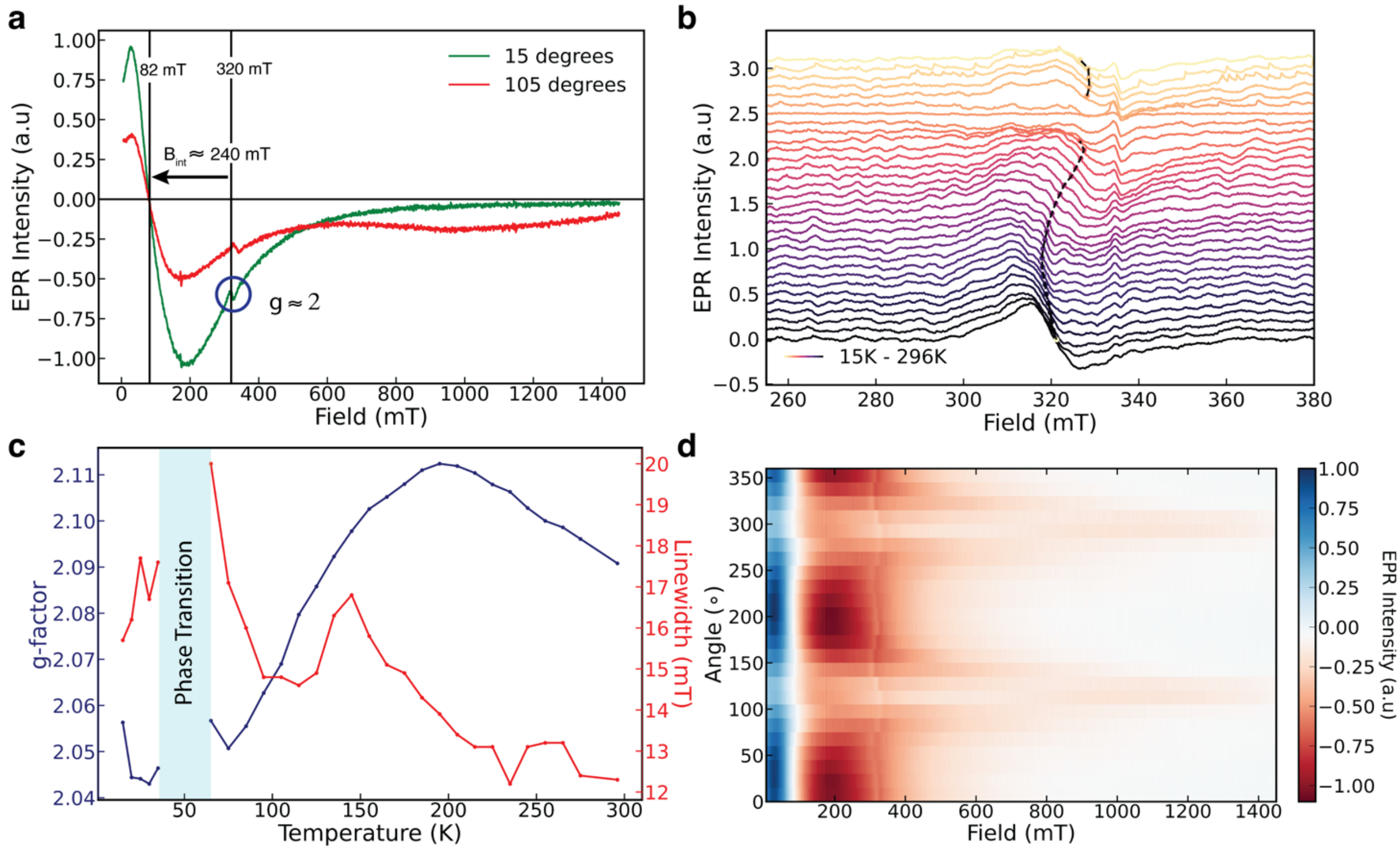

**Figure 2: EPR and FMR local magnetic correlations: a.** cwEPR signal of PNRs in the solid state measured over a wide magnetic field range (at room temperature) shows both the unpaired electron peak at 320 mT sitting atop of a much larger ferromagnetic resonance at 82 mT, suggesting an internal field of 240 mT (green curve; 15°– low field signal: internal field aligned to the magnetic field). The sample (a PNR film made in the inner wall of an EPR tube; see Supplementary Notes 9 and 15) is highly inhomogeneous hence rotation results in a shift in the resonance positions (red curve; 105°– high field signal: internal field anti-aligned to the magnetic field). **b.** Line cuts of the $g \approx 2$ signal along with the zero-intensity crossing (dashed black line) as a function of temperature. **c.** Extracted $g$-factor and linewidth as a function of temperature. We observe a divergence in the linewidth around 40 to 65 K characteristic of a phase transition. **d.** Orientational magnetic anisotropy in the EPR (FMR) signal of the 'inner wall' PNR sample across the full range of rotation angles of tube at room temperature (see Figure S15.13 for the relationship between resonator and EPR tube axes).

On rotating the sample tube in the EPR spectrometer, we discover that the EPR and FMR signals both show a strong orientational dependency around the sample tube long axis, as shown in **Figure 2d** (see Supplementary Note 15 for a diagram of the rotation axes). This behavior arises from the inhomogeneity of the inner wall film in thickness, packing, PNR orientations and size (see Supplementary Note 9).

To shed some light on the internal fields within the EPR film samples, we fit the FMR signal with two purely Lorentzian and one Voigtian component (see Figure S15.5). We find that some PNR domains give rise to net internal fields of ≈225-265 mT and are aligned to the external magnetic field direction, thereby shifting the effective resonance position to lower magnetic fields, at around ≈55-95 mT. We also see certain PNR domains where the effective resonance position is strongly orientation dependent, ranging from 850 mT to 240 mT. This FMR behaviour is reminiscent of reports of magnetism in Ni nanocubes where different sized nanocubes exhibit different internal fields and distributions of internal fields, resulting in a rich orientation dependent behaviour[59]. In these materials, the significant changes in resonance position with changing orientation are indicative of domains that are hard to (re-) magnetize, resulting in low field signals when the net internal field is (close to) aligned to the external magnetic field direction (hard axis), and high field signals when the net internal field is anti-aligned to the external magnetic field direction (easy axis). PNR domains exhibiting an easy and hard axis could explain the FMR signals with weak and strong orientation dependencies respectively. In this case the net internal field of these PNR domains would be estimated to be ≈600 mT considering the minimum and maximum resonance positions achieved *via* sample rotation. However, the inhomogeneous nature of PNR films and broad linewidths of all the FMR signals we observe suggests a wide distribution of internal fields within each domain. Consequently, verifying the presence of such hard and easy axes *e.g.*, by measuring the FMR response after holding the film at high magnetic fields, is challenging (see Supplementary Note 15 for further discussion).

Focusing again on the $g \approx 2$ EPR signal, we see that the $g$-factor and linewidth also vary as a function of sample tube rotation, with the $g$-factor ranging from 2.03 to 2.21 (see Figure S15.3). The $g$-factor for black phosphorous has been calculated to be 2.0 in-plane and 2.14 out-of-plane in the monolayer (1L) limit, with a strong layer number dependence, reaching 2.9 in the bulk out-of-plane limit[60]. The calculated out-of-plane $g$-factor for a 2 layer (2L) black phosphorous is ≈ 2.21. Hence the observed EPR signal, results predominantly from 1L and 2L PNRs, consistent with our measured layer distributions in **Figure 1c** (and Supplementary Note 1)[39], and the spin polarisation is either in-plane or out-of-plane depending on the tube orientation.

From the double integral of the EPR signal in **Figure 2a** (105 degree curve) we can quantify the number of spins that contribute to the signal to be $\sim 4 \times 10^{15}$. Noting the mass of PNRs in the sample (400 μg), distribution of PNR dimensions in solution **Figure 1c**, top panel (and Supplementary Note 1 and 8) and PNR unit cell parameters[61], we can estimate the average spin polarisation length along a PNR to be between 42 and 377 nm (see Supplementary Note 16 for further details and limitations of the calculations).

Having established the signatures of key spin correlations and room temperature magnetism in PNRs, we turn to their optical (semiconducting) properties. While the absorption spectrum of PNRs, as shown in the inset of **Figure 1d**, is broad and featureless, on application of pressure (up to 300 MPa) we observe a large monotonic blue-shift in the band-edge which is 6 times greater than that of phosphorene for the same applied pressure[62] (Supplementary Note 19; the blue-shift is also indicative of monolayers). Electroabsorption measurements reveal further structure to the absorption spectrum, with two transitions at ~410 nm and ~590 nm (charge separated state), and signs of a built-in dipole moment in PNRs (in-keeping with -47 mV surface charge in solution; see Supplementary Note 22). Emission measurements on individual PNRs (Supplementary Note 18) tentatively indicate a band-gap that is inversely proportional to the PNR width and a complex spectrum of likely-excitonic transitions which are inhomogeneously broadened three-fold between the ensemble and single PNRs (Supplementary Note 20). These measurements additionally reveal non-radiative trap activation energies of ~21 meV, radiative/non-radiative rates (from photoluminescence streak camera measurements) and the emission quantum yield of PNRs to vary from 11% to 25% on cooling from 300 to 10 K.

Focussing specifically on the electronic dynamics using optical pump-probe spectroscopy, upon photoexcitation with a 460 nm, 250 fs pulse, we see the emergence of two long lived (nanosecond) excited state absorptive (ESA) bands at 800 nm (ESA 1) and 500 nm (ESA 2) along with an overlapping, short-lived (picosecond) broad photobleaching band at 700 nm (**Figure 3a**). From the LD in a magnetic field (**Figure 1b**, inset), we know that the ribbons likely align with their short axis along the magnetic field direction in solution. In **Figure 3b**, we show the results of performing a polarization resolved pump-probe experiment on the PNRs in solution, with an 800 mT applied external magnetic field in the Faraday geometry. We find that the typical co-polarised horizontal pump - horizontal probe (HH) and vertical pump - vertical probe (VV) signal's degeneracy is lifted and a lab axis is introduced into the experiment upon application of an 800 mT field. The magnetic field is polarised along the H direction in the lab axis which, as expected, results in the higher transient absorption signal due to the increased net linear dichroism along the HH direction (**Figure 1b**). Further, as expected, we see that the alignment effect scales with the applied field strength (see Supplementary Note 21).

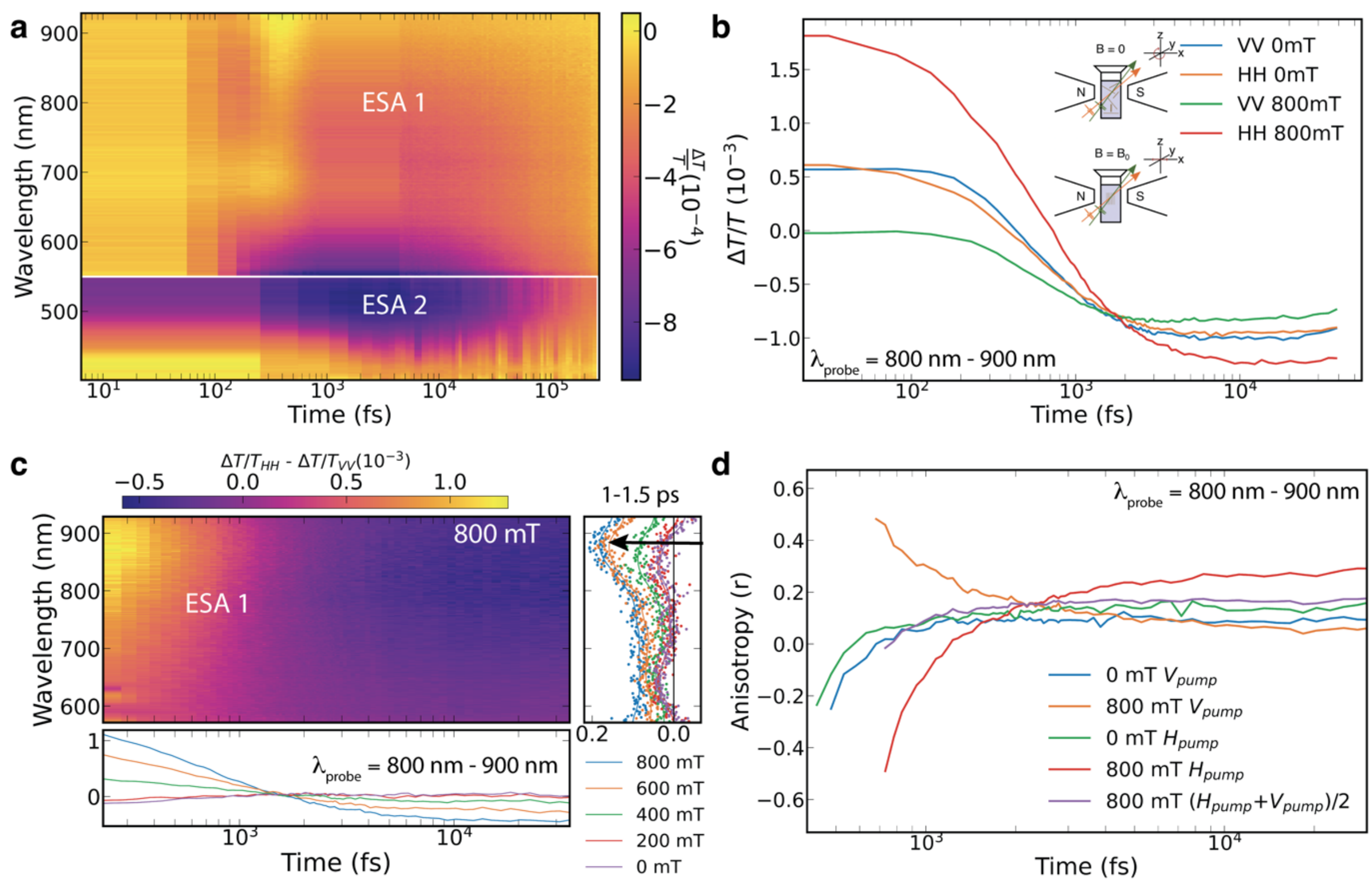

**Figure 3: Photoexcited dipole relaxation dynamics. a.** Broadband transient absorption spectrum of PNRs. Upon photoexcitation the PNRs display two broad excited state absorption (ESA) bands that decay on the order of ns. There appears to be a broad photobleaching signature at 700 nm that overlaps with the ESA bands. **b.** Polarisation and magnetic field resolved decay of ESA 1 kinetics. Upon application of a 800 mT field, the ribbons align in the magnetic field, splitting the degeneracy of the HH and VV polarised experiments. Further, we find a significant reduction in the signal magnitude when photoexciting V polarised (along the PNR long axis). Inset shows the orientation of PNRs with respect to light polarisation and magnetic field. **c.** Magnetic field induced anisotropy obtained by subtracting the spectral response of the unaligned ribbons. This unveils a spectral signature centred at 850 nm (corresponding to ESA 1), the magnitude of which can be strongly modulated with a sub-1T field (right panel). **d.** Decay of polarisation anisotropy ($r$) as a function of magnetic field. In an unpolarised solution under no magnetic field, we find that the anisotropy is initially -0.2 and decays to close to 0 within 1 ps (green and blue), indicating the photoexcited state is perpendicular to the probed state after 1 ps (the differences in the V and H polarised experiments at 0 mT demonstrate the setup's resolution). Upon application of the 800 mT field however, photoexciting V polarised (PNR long axis dipole, orange curve), we find that the anisotropy retains its sign (indicating the photoexcited state is the same orientation as the probed one) whereas when photoexciting H polarised (PNR short axis dipole, red curve), the anisotropy flips sign, indicating a dipole flip, as in the unpolarised case, with the average converging to the unpolarised system, as expected.

We leverage this magnetic alignment effect to isolate the spectral features of the magnetically aligned ribbons (as opposed to optical signatures of a non-magnetic sub-population) by subtracting the HH and VV polarised transient absorption maps at different field strengths, as any unaligned ribbons would have identical HH and VV polarised transient absorption maps, yielding no difference signal. As seen in **Figure 3c**, we find that the magnitude and kinetics of the photoinduced absorption band (ESA 1) is most strongly perturbed by the external field. We can therefore confidently assign ESA 1 to that of the magnetically-active ribbons.

Interestingly, ESA 1 (**Figure 3a**) forms quickly (sub-1 ps) upon the disappearance of the two positive photobleaching-like spectral features (which could also arise from stimulated emission as they are aligned with the photoluminescence bands in Supplementary Note 20). In order to understand this and track the evolution of the photoexcited dipole's relaxation dynamics in the magnetically active ribbons, we monitor the transient absorption anisotropy in the ESA 1 band (800-900 nm) defined as

$$r = \frac{\frac{\Delta T}{T}_{\parallel} - \frac{\Delta T}{T}_{\perp}}{\frac{\Delta T}{T}_{\parallel} + 2\frac{\Delta T}{T}_{\perp}} ,$$

where $\frac{\Delta T}{T}_{\parallel}$ and $\frac{\Delta T}{T}_{\perp}$ are the transient absorption signal with the probe polarisation parallel and perpendicular to the pump polarisation[63]. We find that, in the absence of an external field, at 200 fs the anisotropy is -0.2 (blue/green traces), signifying that we are likely probing the dipole polarised orthogonally to the main photoexcited transition. Over 1 ps, the anisotropy relaxes to a value of about 0.05, signifying that there is energetic transfer between the dipoles as the timescale is faster than the ns-orientational relaxation dynamics typically associated with molecular anisotropy. As expected, we demonstrate the H and V pumped anisotropies are degenerate, demonstrating the instrument's anisotropy resolution (**Figure 3d**, blue and green).

Upon application of the 800 mT field, due to the macroscopic alignment of the ribbons, we are able to increase our dipole selectively and pump the short-axis polarised (H) and long-axis polarised (V) dipole. It is worth noting that as the ribbons are now aligned, the sign of the anisotropy is the key feature of interest, as the value of the anisotropy will not be consistent with the isotropic molecular distribution assumption standard to such measurements[64]. We find that when pumping the long-axis polarised (V) dipole (**Figure 3d**, orange), the anisotropy does not change sign, indicating that the energy remains in this state, whereas when we pump the short-axis polarised (H) dipole, the sign flips (red), just like in the case of the unpolarised ribbons (blue). Further, the fact that we can construct the unaligned PNR anisotropy as an average of pumping the 800 mT aligned long-axis polarised (V) and short-axis polarised (H) dipoles (**Figure 3d**, purple) confirms that these are the main two dipoles involved in the optical response of the system. Taken together with the larger magnitude of the transient absorption signal (**Figure 3b**) of the H pumped system, this suggests that the photoexcitation is initially predominantly polarised along the short axis and transfers to an orthogonal dipole polarised along the PNRs long axis on a sub-1 ps timescale. We remark that single PNR (polarisation resolved) transient absorption microscopy measurements reveal a distribution (~0.75 ps standard deviation) in the ESA 1 lifetimes, with the broadness of the ESA 1 band also shown to be three-times lower at the single PNR level (see Supplementary Note 18 and Supplementary Note 20 where we show the same behaviour for the emission). We hence highlight that the timescales above are an ensemble average.

GW-BSE calculations of the optical spectrum of zigzag PNRs (see also footnote[a]) report that the lowest excited state is a dark exciton which is polarised along the ribbons long axis[37]. This is consistent with the sub-ns emission lifetimes and relatively low emission quantum yield we measure in the PNRs (see Supplementary Note 20). Taking this with the data in **Figure 3** we propose that the primary photoexcitation mechanism is therefore that the initial excitation dipole polarised along the ribbon short axis decays to this long axis polarised dark state on sub-1 ps timescales. This is consistent with the fast disappearance of the photo bleach signal corresponding to the initial state in the transient absorption spectra in **Figure 3a**, with the formation of long lived ESAs likely from the dark state.

While calculations have reported that the first bright exciton in PNRs, like those we study experimentally, has significant density along the ribbon edges, it is unclear whether this lower energy dark state that the photoexcitation decays to is also localised along the ribbon edges[37].

This is critical for any potential excited state magneto-optical coupling in PNRs, as most of the photoexcitation lifetime appears to be spent in this state. To study this we exploit the presence of special localised phonon modes on the PNR edges[65].

The spontaneous Raman spectrum of the PNRs (**Figure 4a**, blue curve) is dominated by $A^1{}_g$, $B^2{}_g$ and $A^2{}_g$ symmetry modes, with additional bulk-symmetry forbidden $B^1{}_g$ and $B_{3g}{}^1$ between 190 and 250 $cm^{-1}$. All of these mode are also observed in the Raman spectrum of black phosphorous (BP)[66] and we demonstrate (as also previously-reported) that the $B^1{}_g$ and $B_{3g}{}^1$ modes specifically lie along the zigzag edge by performing Raman imaging of BP (**Figure 4a**, inset)[65]. In order to study the mode coupling to the excited state energy surface, we perform impulsive vibrational spectroscopy (IVS) using broadband compressed 13-fs pulses centred at 550 nm. We find that when resonantly photoexciting the PNRs, the photoexcited state is strongly coupled to the $B_{3g}{}^1$ edge phonon mode (**Figure 4a**, red curve).

Control experiments which involve performing off-resonant IVS on the PNRs and DMF solvent confirm that the $B_{3g}{}^1$ edge phonon mode is indeed dominantly coupled to the excited state potential energy surface (see Supplementary Note 23). Furthermore, as seen in **Figure 4b**, we find that this edge phonon mode is most strongly coupled to ESA 1, the transition characteristic of the magnetically active ribbons. The selective coupling of the excited state to the edge mode as opposed to the bulk phonon modes suggests that the electron density of the photoexcitation is along the edges of the PNRs. In the solution phase, we observe that the excited state IVS of the $B_{3g}{}^1$ out-of-plane edge mode is present when studying the unaligned (0 mT) ribbon solution only in the perpendicular pump-probe configuration, *i.e.* HV or VH (see Supplementary Note 23) and not the parallel configurations. In other words – one needs to probe a dipole perpendicular to the main photoexcitation (that we ascribe to the transition polarised along the PNRs long axis) in order to observe the excited state moving along the $B_{3g}{}^1$ vibrational coordinates at the edge of the PNRs. Interestingly, modelling the temperature evolution of the photoluminescence/absorption spectra, we also find that this single edge mode (220 $cm^{-1} \equiv 317K$) appears to have the dominant role in electron-phonon coupling, *i.e.*, charge relaxation, in PNRs (see Supplementary Note 20).

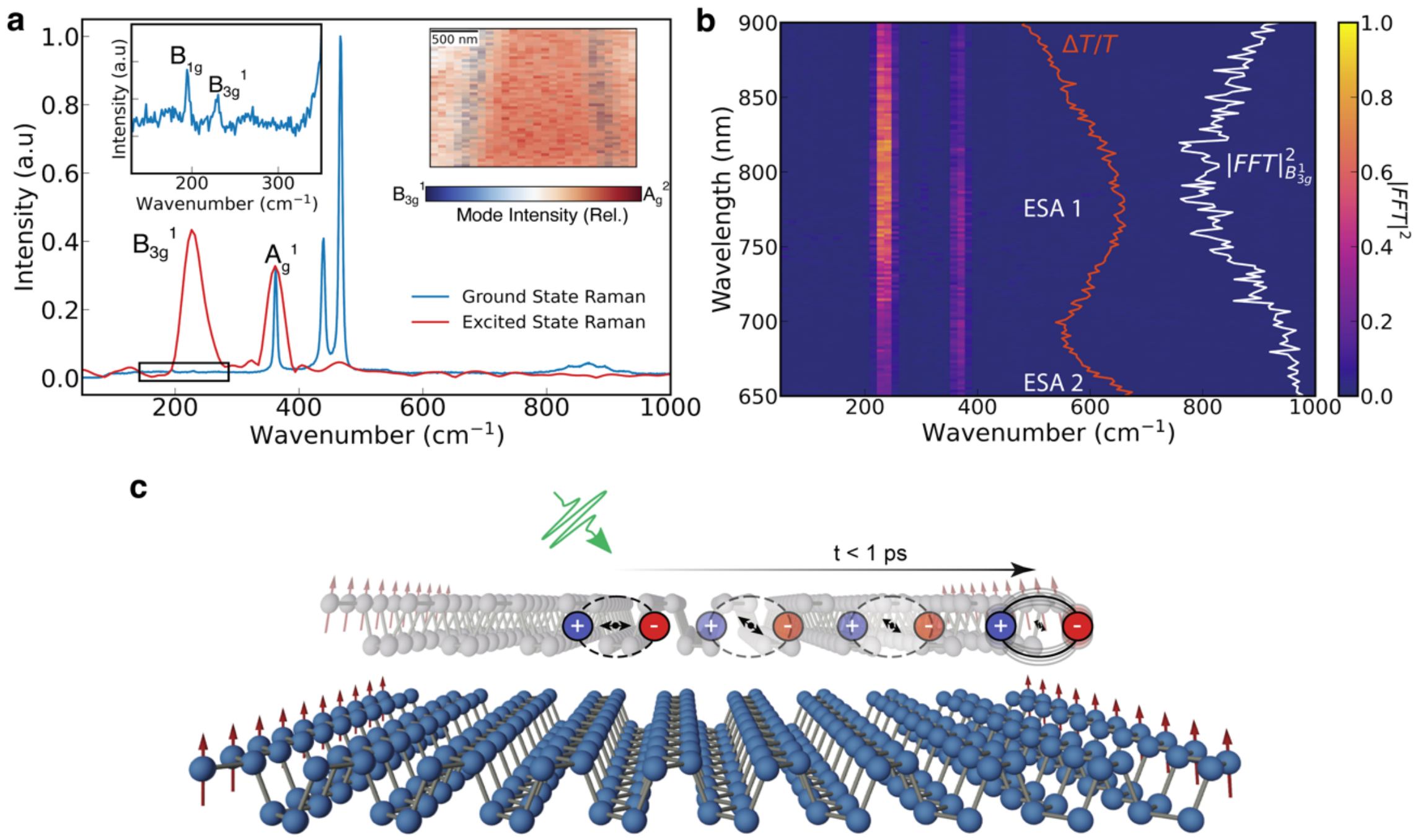

**Figure 4: Coupling of the photoexcitation to the magnetic edges a.** The excited state (impulsive, time domain) Raman spectrum of the PNRs (red) shows a strong coupling to the $B_{3g}{}^{1}$ mode compared to the ground state spectrum (blue), indicating that the photoexcitation is delocalised along the edges of the ribbon. Inset: (Left) Zoom into the low frequency region of the ground state Raman spectrum. (Right) Raman imaging of a black phosphorous flake shows that, as previously reported[65], the $B_{3g}{}^{1}$ mode is localised on the edge. We note that the (broad) linewidth in the IVS experiments is as a result of the instrument resolution/Fourier transforming of the time domain spectrum, not inhomogeneous/lifetime broadening. **b.** Wavelength resolved, impulsive, time-domain Raman spectrum of the photoexcited PNRs reveals that the $B_{3g}{}^{1}$ mode is most strongly coupled (white curve) to ESA 1 (red curve) which was also most strongly modulated by the external magnetic field. **c.** Schematic of the photoexcited dipole dynamics, where the excited state is initially polarized along the PNR short axis and, on sub-1 ps timescales, relaxes to a dipole polarized along the long axis with significant density along the ribbon edge. Here, the excitation is coupled to a symmetry forbidden edge phonon mode and it is at the edge where the magnetism is likely present.

In summary, we have evidenced *via* SQUID magnetometry and cwEPR, macroscopic magnetism at room temperature in films of PNRs, arising from (likely) the ribbon edges. In solution a giant magnetic anisotropy means PNRs can be readily aligned with their short axis along the field direction. As shown in **Figure 4c** and description of the various electronic states above, on photoexcitation energy rapidly migrates from a bright- to dark-photoexcited state that is localised to the PNR edge and strongly coupled to a symmetry-forbidden edge phonon mode where the magnetism likely exists. Our findings open up extensive opportunities for further understanding the magnetic properties of PNRs (spin coherence times, influence of ribbon width/strain, role of thickness, *etc*), building of proposed single nanoribbon devices[67–71] and for more disparate applications (non-linear optical elements[72–74], nanorobotics[49,75,76], magnetic drug delivery[77] and in photocatalysis[57]). Important questions are raised about the exact microscopic arrangement of atoms/spins at the ribbon edge and the in/out-of-plane couplings and couplings between layers. However, our STM measurements (Supplementary Note 5) and polarised impulsive Raman measurements (see also Supplementary Note 2) provide a first step towards this. While we show PNRs to be magnetic and optically active semiconductors, the spin-photon/phonon coupling mechanism also requires further exploration. Nonetheless, the optically active nature of PNR edges provides a promising a route for nanoscale magnetic-photon interfacing *via* excitation of either the electronic states involved in ESA 1 or direct excitation of edge phonons.

## Methods

### PNR Synthesis

PNRs were synthesised as reported in Watts *et al.*[30]. In short, in an argon glovebox (<0.1 ppm $O_2$, <0.1 ppm $H_2O$) black phosphorus (2D semiconductors Ltd and Smart Elements) was ground *via* pestle and mortar into ca. 1 mm flakes and 124 mg (in a typical synthesis) was transferred to a glass tube fitted with a metal Swagelok valve, alongside 3.5 mg freshly cut lithium metal (Sigma Aldrich, SA, 99% rod). The tube was evacuated (ca. $10^{-7}$ mbar) and cooled to -50 °C and ammonia gas (SA, 99.95%, precleaned by condensation over excess lithium metal) was condensed to submerge the lithium and phosphorus. The solution immediately turned dark blue from the formation of lithium-ammonia solution, and slowly turned orange over 16 h. The ammonia was then evaporated and the orange $LiP_8$ salt was dried under vacuum (ca. $10^{-7}$ mbar) at room temperature. The $LiP_8$ (10 mg) was then placed in 10 mL of either NMP (SA, 99.5% anhydrous) or DMF (SA, 99.9% anhydrous) which had been dried with 4 Å molecular sieves for 1 week. The mixtures were bath sonicated for 30 min, before centrifuging (100*g*, 10 min) and decanting in a glovebox to give solutions of PNRs.

### Single PNR Atomic Force Microscopy

PNR samples were prepared *via* drop-casting from solution onto freshly cleaved graphite (HOPG) substrates as discussed in Watts *et al.*[30]. High-resolution topography maps were then collected using contact-mode high speed atomic force microscopy (HS-AFM; Bristol Nano Dynamics Ltd, UK) with silicon nitride microcantilevers (MSNL-10, Bruker) with nominal tip radii of 2 nm. To ensure that the images obtained featured tip-limited lateral resolution scan sizes of 0.8 $\mu$m × 0.5 $\mu$m were used, corresponding to pixel sizes of ~1 nm × 1 nm.

HS-AFM images containing ribbons selected for analysis were input into custom LabVIEW image analysis software (National Instruments) that isolated the ribbon from the background in an automated fashion before generating histograms of ribbon height and width. In brief, the algorithm consisted of: a local thresholding step to identify the ribbon; a mask step to separate pixels corresponding to the background from the ribbon; an erosion step to remove the edge pixels where the AFM tip traverses the slope between the background and ribbon; calculation of height histograms for both the background and ribbon pixels; and finally Gaussian fits to the histograms to recover the mean height of each and enable calculation of the height of the ribbon with respect to the background.

The width of the ribbon was analysed in parallel using the same algorithm. After the masking step the central axis of the isolated ribbon was determined by fitting cubic splines to the two edges of the ribbon and calculating the mean of these two edges. Then, the width of the ribbon was determined by calculating the distance to the central axis from each edge. Measurements of the width of the ribbon along its entire length were then input into a histogram and fit with a Gaussian distribution to identify the mean width of the PNR.

### Magnetic Linear Dichroism and Birefringence

The degree of magnetic alignment is measured optically, through the magnetic field-induced linear dichroism (LD) and birefringence (LB), using green and red Helium Neon lasers (wavelengths 543 nm and 632.5, respectively). The solution samples were contained in an optical cuvette (thickness 5 mm or 10 mm) positioned within a temperature-controlled environment at 20.0 ± 0.1 °C in a 33 T Florida-Bitter electromagnet or a Varian V-3900 2 T magnet. The LD and LB signals were measured using standard polarization modulation techniques using a photo-elastic modulator[78].

<u>SQUID Magnetometry</u>

Magnetisation measurements were obtained in a Quantum Design Magnetic Properties Measurement System (MPMS 3) using a Superconducting Quantum Interference Device (SQUID) magnetometer. Measurements were performed down to 1.8 K and up to 350 K, in various applied magnetic field strengths up to 7 T. Data for two isothermal magnetic field scans of PNRs out of solution are shown in **Figure 1d**, and two magnetisation-temperature curves of PNRs are displayed in the inset of **Figure 1d**. Each magnetisation-temperature curve was measured starting at 1.8 K and increasing the temperature in a constant probing field of 50 mT. Prior to beginning any new magnetic field or temperature scan the system was taken to 300 K and the magnet was reset to remove any stray flux from the SQUID. Samples were secured on an MPMS quartz sample holder using GE varnish, and care was taken to ensure that they were not touched by any magnetic material throughout the mounting and loading procedure. For each measurement, several dc magnetisation measurements were averaged, providing a reliable measure of the bulk magnetisation of the PNR samples.

For measuring the PNRs in the DMF and NMP solutions (Figure S10.2) 10 μL were pipetted into a plastic capsule. The capsule was fixed to a quartz rod using GE varnish, and the presence of PNRs increased the magnetisation by two orders of magnitude when compared to the GE and capsule by itself. The dc moment was averaged over multiple scans, and between each experimental run the MPMS was brought to 300 K and the magnet was reset to remove any remaining flux in the SQUID. Oxygenation of the PNR was performed by bubbling air through PNRs in the plastic capsule for ~20 mins while ensuring the sample it did not come into contact with any metallic substances. See also Supplementary Notes 10, 11 and 24 for further details.

<u>Electron Paramagnetic Resonance</u>

For the cwEPR experiments, 100 $\mu$L of PNR in DMF solution was filled into a 3.9 mm outer diameter (2.9 mm inner diameter) quartz EPR tube, inside a nitrogen glovebox. The sample tube was attached to a custom adaptor and transferred to a pumping station outside the glovebox. The custom adaptor keeps the sample in the inert glovebox environment. The solution was evaporated under vacuum (using a pumping station), resulting in a film on the inner walls of the EPR tube. The procedure was repeated four times, to achieve a thicker film, in total 400 $\mu$L was evaporated. The inner wall sample was then left to pump to a pressure of $6\times10^{-4}$ mbar and flame-sealed.

*Narrow Magnetic Field Scan - Rotation and Temperature Series (Setup 1):*

The cwEPR spectra were recorded at X-band (≈9.4 GHz) using a laboratory-built EPR spectrometer (all narrow scan cwEPR spectra were rescaled to 9.4 GHz). The setup consists of a Bruker ER 041 MR microwave bridge together with an ER 048 R microwave controller and an AEG electromagnet together with a Bruker BH 15 Hall effect field controller. The magnetic field was additionally monitored with a Bruker ER 035 M NMR Gaussmeter. The resonator used was a Bruker ER 4122-SHQE resonator. The static magnetic field was modulated at 100 kHz and lock-in detection was carried out using a Stanford Research SR810 lock-in amplifier in combination with a Wangine WPA-120 audio amplifier. An ESR 900 helium flow cryostat together with a ITC503 temperature controller (Oxford Instruments, UK) was used for low temperature measurements. The spectra were acquired at a frequency of ≈9.4 GHz with a microwave power of 7.96 mW and 1 mT modulation amplitude. The magnetic field was calibrated using a standard N@C60 sample with a known *g*-factor.

The narrow magnetic field scans presented in the main text (**Figure 2c**) have had the FMR signal (slope) removed from them using an appropriate polynomial fit. See Supplementary Note 15 for further details.

*Wide Magnetic Field Scan - Rotation Study (Setup 2):*

The cwEPR spectra were recorded at X-band (≈9.55 GHz) using a laboratory-built EPR spectrometer (All wide scan cwEPR spectra were rescaled to 9.55 GHz). The setup consists of a Bruker ER 046 XK - T microwave bridge together with an ER 048 R microwave controller and a Varian electromagnet together with a Bruker ER 032M Hall effect field controller. The resonator used was a Bruker MD5 dielectric ring resonator. The static magnetic field was modulated at 99 kHz and lock-in detection was carried out using a Stanford Research SR830 lock-in amplifier in combination with a Wavetek 50 MHz function generator model 80. The wide magnetic field measurement was carried out at 296 K. The spectra were acquired at a frequency of ≈9.55 GHz with a microwave power of 7.96 μW and 0.4 mT modulation amplitude.

*Wide Magnetic Field Scan without low-temperature insert - Rotation Series (Setup 3):*

The wide magnetic field scan without a low-temperature insert, was carried out on a Bruker ElexSys E580 spectrometer, with a Bruker ER 4122-SHQE resonator. The SHQE cavity is the same type of cavity as used in the narrow field range setup 'Setup 1'. However, this time we use the cavity without the low-temperature Dewar insert, as this insert can have background signals. The reason for using the Bruker ElexSys E580 is that this allows us to do the full magnetic field sweep using the SHQE resonator. The spectra were acquired at a frequency of ≈9.85 GHz with a microwave power of 8.05 μW and 0.4 mT modulation amplitude.

Scanning Tunnelling Microscopy (STM)

All STM experiments were performed on a commercial Omicron LT-STM at 4.2 K using PtIr STM tips. Samples were prepared by aerosolizing PNRs dispersed in NMP onto a HOPG substrate to minimize aggregation on the surface[79]. The deposition parameters and the concentration/density of PNRs on the surface were optimized using ambient atomic force microscopy imaging. For UHV-STM experiments PNRs were aerosolized onto a freshly cleaved HOPG substrate using an Iwata Custom Micron CM-C airbrush. The coated substrate was immediately transferred into the vacuum chamber and annealed for >1 h at 150 ˚C before transferring the substrate to the STM sample stage (extended annealing or heating > 150 ˚C led to partial degradation of PNRs). Supplementary Note 5 contains further details on the STM imaging.

Transient Absorption Spectroscopy

The transient absorption measurements from 550 – 930nm were performed using a homebuilt setup around a Yb:KGW amplifier laser (1030 nm, 38 kHz, 15 W, Pharos, LightConversion). The probe pulse was a chirped seeded white light continuum created using a 4 mm YAG crystal that spanned from 500 nm to 950 nm. For the source of the pump pulse (~200 fs) a commercial optical parametric amplifier (OPA) OPHEUS ONE (Light Conversion) was used.

The transient absorption measurements from 400 – 550 nm were performed using a homebuilt setup around a Ti-Sapphire (800 nm, 1 kHz, Spectra-Physics, Solstice Ace). The probe pulse was a chirped seeded white light continuum created using a 4 mm $CaF_2$ crystal that spanned from 400 nm to 600 nm. For the source of the pump pulse (~100 fs), the fundamental of the laser was doubled in a β-barium borate crystal.

The sub-1 T magnetic field was generated using an electromagnet from GMW Model 3470 with 1 cm distance between cylindrical poles and the field strength calibrated with a Gaussmeter.

Impulsive Raman Spectroscopy

Femtosecond time domain Raman spectroscopy measurements were performed using a homebuilt setup around a Yb:KGW amplifier laser (1030 nm, 38 kHz, 15 W, Pharos, LightConversion). The probe pulse was a chirped seeded white light continuum created using a 4 mm YAG crystal that spanned from 500 nm to 950 nm. The pump pulse for the resonant experiment was created using a non-collinear optical parametric amplifier (NOPA) where the 1030 nm seeded a white light continuum stage in sapphire which was subsequently amplified with the third harmonic of the 1030 nm laser in a β-barium borate crystal to create a broad pulse centred at 550 nm. The pump pulse for the off-resonant experiment was created using a non-collinear optical parametric amplifier (NOPA) where the 1030 nm seeded a white light continuum stage in a YAG crystal which was subsequently amplified with the third harmonic of the 1030 nm laser in a β-barium borate crystal to create a broad pulse centred at 750 nm. Both pulses were compressed using a chirped mirror and wedge prism (Layerterc) combination to a temporal duration of under 15 fs. Compression was determined by second-harmonic generation frequency-resolved optical gating (SHG-FROG; upper limit) and further confirmed by reference measurements on acetonitrile where the 2,200 $cm^{-1}$ mode could be resolved. The probe white light was delayed using a computer-controlled piezoelectric translation stage (Physik Instrumente), and a sequence of probe pulses with and without pump was generated using a chopper wheel (Thorlabs) on the pump beam. The average fluence of the pump 10 μJ /$cm^2$.

Transient Absorption Microscopy

Pulses were delivered by a Yb:KGW amplifier (Pharos, LightConversion, 1030 nm, 5 W, 200 kHz) that seeded two broadband white light (WL) stages. The probe WL was generated in a 3 mm YAG crystal and adjusted to cover the wavelength range from 650–950 nm by a fused-silica prism-based spectral filter. In contrast, the pump WL was generated in a 3 mm sapphire crystal to extend the WL in the high frequency to 500 nm, with pulse short-pass filtered at 650 nm (Thorlabs, FESH650). The pump pulses were focussed onto the sample using a single-lens oil immersion objective (100×, numerical aperture 1.1 NA) to a diffraction-limited spot of ~270 nm (FWHM, full bandwidth). In contrast, the counter-propagating probe pulses were loosely focused onto the sample by a concave mirror (FWHM ~15 μm). A set of third-order corrected chirped mirrors (pump WL: Layertec, probe WL: Venteon) in combination with a pair of fused silica wedge prisms (Layertec) compressed the pulses to sub-15 fs at the sample. The transmitted probe light was collected by the same objective used to focus the pump pulses. The probe was then relayed to a spectrometer consisting of a slit in the intermediate image plane and a F2 prism to disperse the light perpendicular to the slit. This allowed us to access spectrally-dispersed transient absorption microscopy images of the PNRs over a selected region. The differential nature of the imaging was achieved by modulating the pump beam at 45 Hz by a mechanical chopper. All recording was performed with an EMCCD camera (Qimaging Rolera Thunder, Photonmetrics). The axial focus position was maintained by an additional auto-focus line based on total internal reflection of a 405 nm continuous wave laser beam.

Ensemble photoluminescence measurements

*Steady-state photoluminescence:*

PL experiments are performed using a home-made confocal-like setup using a large working-distance microscope objective (NA ≈ 0.55) to focus light on the nano-ribbons (NRs) and collect the emission in a reflection configuration. In the studies PNRs are dispersed (dropcasting method) on glass slides (thickness ≈100 μm) and glued (with silver lacquer) on the cold finger of a cryostat designed for thermal expansion compensation (from Oxford Instruments). The PL signal is analyzed using to a 75 cm focal length spectrometer (Acton sp2750i, Princeton Instruments) itself coupled to a nitrogen cooled CCD (Spec10, Princeton Instruments), a combination that leads to a ≈100 μeV energy resolution, well beyond the resolution required to address the broad PNRs PL components. To minimize the scattered light, the excitation is tuned to 416 nm (SHG of a Ti:Sapphire laser, pulse width of ≈2 ps, 80 MHz repetition rate) and a dichroic filter (Semrock FF01-430/LP-25, 50% cut-off at 437 nm) is placed in the detection path. The PL polarization is analyzed using a classical scheme: a motorized half-wave plate, positioned upstream along the detection beam path, allows us to rotate the polarization of the PNR emission that is further analyzed using a polarizer (Glan-Taylor, calcite) placed in front of the spectrometer slit (with its direction set parallel to the grating grooves in order to enhance the CCD response).

*Time resolved photoluminescence:*

The spectro-temporal PL maps are measured using a streak-camera synchronized with the high repetition rate Ti:Sapphire laser (C5680 model from Hamamatsu incorporating a M5675 synchroscan unit). All measurements are carried out in the confocal configuration described above, the camera being directly coupled to the Acton spectrometer, using a deflection mirror and imaging the NRs spectra on the entrance slit of the camera. The response function of the camera is measured as the response to laser excitation with an IRF curve that demonstrates a time resolution of ≈22 ps across the detection range to the setup. The same cryostat and mounting as for steady-state PL measurements were used for the time-resolved measurements.

Single PNR photoluminescence imaging and spectroscopy

Spinning disk confocal photoluminescence (PL) measurements were performed with a commercial Nikon X-Light V2 microscope, with a 100× 1.52 NA oil immersion objective.

Excitation in all cases was at 390 nm with nominal laser powers (at the laser output) between 20-50 mW. Imaging was performed with a 40 μm confocal pinhole and iXon 897 EMCCD Camera (Andor). The lateral resolution in such measurements is around 120 nm. For the measurement of PL spectra from individual PNRs, the microscope emission outcoupling-arm was modified to pass the PL to a spectrograph (Kymeria 193i; 600 lines per mm grating; 600 nm blaze) with the emission once again measured by an iXon 897 EMCCD. Typical accumulation times of 1-2 mins were used to collect spectra from the PNRs with a collection area on the sample of 0.1 $\mu m^2$.

<u>Absorption spectroscopy</u>

*Temperature dependent PNR absorption*

An Agilent Cary 6000i UV–vis–NIR spectrophotometer with blank substrate correction was used. 400 μl of PNR solution was dropcast onto fused silica substrates and placed in a continuous-flow cryostat (Oxford Instruments Optistat CF-V) under a continuous inert atmosphere. Samples were cooled to 6 K with the temperature dependent absorption taken on heating.

*Individual PNR absorption spectroscopy*

Absorption spectroscopy of individual (large, >300 nm width) PNRs was performed on a customised Zeiss Axio microscope with illumination provided by a halogen lamp (Zeiss HAL100). Transmitted light was collected using a 50×/0.4 objective (Nikon, T Plan SLWD) and spatially filtered using a 100 μm-diameter optical fibre (Avantes FC-UV100-2-SR) mounted in confocal configuration and connected to a spectrometer (Avantes AvaSpec-HS2048). PNR samples were prepared *via* drop-casting from solution onto cleaned (Acetone/IPA) 0.17 mm thick glass slides.

<u>Pressure dependent absorption</u>

In order to study the pressure dependence of the PNRs, transmittance spectra are measured with a LAMBDA 750 UV/Vis/NIR Spectrophotometer (Perkin Elmer). The PNR solutions are dried in an inert atmosphere and then placed inside a high-pressure cell (ISS Inc.) filled with an inert liquid, Fluorinert FC-72 (3M). Hydrostatic pressure is generated through a pressurizing liquid using a manual pump. Prior to using, the liquid is degassed in a Schlenk line to remove oxygen which cause, from 300 MPa onwards, scattering of a fraction of light and therefore a reduction of the transmitted signal from the sample. The pressure is applied from ambient pressure to 300 MPa in steps of 50 MPa. Before the measurement, we wait 7 minutes for equilibration of the material under pressure. We estimate an error of the pressure reading to be 20 MPa.

<u>Raman measurements</u>

*Temperature dependent Raman spectroscopy*

Raman spectra were measured as a function of temperature from 4 K to 300 K. Raman measurements were conducted by backscattering (T64000, Horiba) a CW diode line (532 nm, 1 mW). Spectra were collected at >200 $cm^{-1}$, where the CCD detector (Horiba Synapse Open-Electrode) has a monotonically increasing quantum efficiency of 0.43 – 0.50. Acquisitions employed a 100× optical objective and used minimal laser intensity to avoid sample degradation.

*Raman imaging*

For Raman imaging a standard layout of an epi-detected Raman microscope was used. A pump laser beam (wavelength = 532 nm, Coherent Mira) was spectrally cleaned up by a bandpass filter (FLH05532-4, Thorlabs), and its beam width was expanded to 7.2 mm before entering a home-built inverted microscope. Additional waveplates (half-waveplate and quarter-waveplate for 532 nm, Foctek Photonics) pre-compensated the ellipticity introduced by the dichroic filter (F38-532_T1, AHF) and also generated circularly polarised light. We used high numerical aperture (NA) oil-immersion objectives (Nikon 60X/1.4 N.A. oil) to ensure high-resolution imaging and increase collection efficiency. The pump power before the objective was 30 mW, a power level that ensured no degradation of samples. The samples were scanned with galvanometric mirrors (Thorlabs). The Raman inelastic backscattered light was collected by the same objective and focused with the microscope tube lens either onto the slit of a spectrometer (Andor, Shamrock 303i, grating 300 l/mm; the slit also acts an effective pinhole for confocal detection). The spectrometer is equipped with a high-sensitivity charge-coupled camera (Andor, iXon 897). All images presented were taken with integration times/pixel in the 500 ms range. Recording of data was performed by a custom Matlab program. For Raman imaging, black phosphorous flakes (300-600 nm thickness; 2D semiconductors Ltd) were mechanically exfoliated inside a nitrogen glovebox and transferred onto a Si substrate. A 0.15 mm thick coverslip was placed over the flakes and sealed with epoxy glue to act as an encapsulant. The polarisation of the pump and Raman light was not strongly controlled, but simply adjusted to maximise the respective signals.

**Acknowledgements:** R.P and A.A thank Russell P. Cowburn (University of Cambridge) for critical reading of the text and assistance with interpretation of the SQUID data, S.S Rao (University of Texas at El Paso) and S.V Bhat (Indian Institute of Science) for discussion of the EPR signals, and Matthew J. Cliffe (University of Nottingham), Cheng Liu (University of Cambridge) and Siân Dutton (University of Cambridge) for discussion/assistance with analysis of the SQUID data. The authours also thank Thierry Barisein, Laurent Legrand and Victor Guilloux (Sorbonne Université) for performing the low temperatire steady-state and time-resolved emission measurements. The authors thank Henning Sirringhaus (University of Cambridge) for use of temperature dependent Raman facilities. R.P. and A.A. thank Akshay Rao (University of Cambridge) for overall support and guidance with the work.

**Funding:** A.A. acknowledges funding from the Gates Cambridge Trust as well as support from the Winton Programme for the Physics of Sustainability including support for a 'Cambridge-Berkley Exchange'. S.F. acknowledges funding from the Studienstiftung des deutschen Volkes and the Engineering and Physical Sciences Research Council (EPSRC UK) *via* an EPSRC Doctoral Prize Fellowship. R.P. acknowledges Clare College, Cambridge for funding *via* a Junior Research Fellowship. We acknowledge financial support from the EPSRC *via* grants EP/M006360/1 and EP/W017091/1 and the Winton Program for the Physics of Sustainability. This work was supported by HFML-RU/NWO-I, member(s) of the European Magnetic Field Laboratory (EMFL) and by EPSRC (UK) *via* its membership to the EMFL (grant no. EP/N01085X/1).

**Competing interests**: The authors declare that they have no competing interests.

**Data and materials availability**: The data that support the plots within this paper and other findings of this study are available at the University of Cambridge Repository (https://doi.org/XXXXX).

**Code availability**: The code used within this study is available at the University of Cambridge Repository (https://doi.org/XXXXX).

**Author Contributions:** A.A. designed and performed optical and magnetic measurements, analysed and interpreted the data and wrote the manuscript. A.J.C., R.R.C.S., T.J.M. and E.S.Y.A. prepared nanoribbon solutions. N.A.P. performed and interpreted the EPR experiments under the supervision of J.B.. N.J.M.P. and A.E. assisted with SQUID experiments. T.G.P. performed micro-absorption experiments. L.P. performed the high-speed AFM measurements and analysed the results. S.F. attempted emission measurements and R.C. performed Raman measurements. L.M. performed pressure dependent absorption measurements. A.C. assisted A.A. in STM measurements under the supervision of F.F.. M.E.S. and S.K. performed magnetic birefringence and magneto-linear dichroism experiments under the supervision of P.C.M.C who interpreted the results. H.B.A supervised the Raman imaging. C.A.H. supervised the production of nanoribbons, interpreted the data and wrote the manuscript. R.H.F. supervised the work of A.A.. R.P. designed and supervised the project, performed optical measurements, interpreted data and wrote the manuscript.